\documentclass{revtex4}   % style for Physical Review B and AJP are similar
\usepackage{amsmath}    % need for subequations
\usepackage{amssymb}
\usepackage{graphicx}   % for figures

\usepackage[latin1]{inputenc}    % pour les accents

\usepackage{float}  % pour que les images soient inclusent dans le texte

\begin{document}

\title{Revisiting the Schrödinger Probability Current}
\author{Michel Gondran}
 %\altaffiliation[Also at ]{home.}  %  optional
 \affiliation{EDF, Research and Development, 92140 Clamart}
 \email{michel.gondran@edf.fr}   %optional
\author{Alexandre Gondran}
 %\altaffiliation[Also at ]{home.}  %  optional
 \affiliation{Paris VI University, Ecole Doctorale "Ondes et Matière"}
 \email{alexandre.gondran@etu.umpc.fr}   %optional

\begin{abstract}
We revisit the definition of the probability current for the
Schrödinger equation. First, we prove that the Dirac probability
currents of stationary wave functions of the hydrogen atom and of
the isotrop harmonic oscillator are not nil and correspond to a
circular rotation of the probability. Then, we recall how it is
necessary to add to classical Pauli and Schrödinger currents, an
additional spin-dependant current, the Gordan current.

Consequently, we get a circular probability current in the
Schrödinger approximation for the hydrogen atom and the isotrop
harmonic oscillator.
\end{abstract}

\maketitle

%%%%%%%%%%%%%%%%%%%%%%%%%%%%%%%%%%%%%%%%%%%%%%%%%%%%%%%%%%%%%%%%%%%%%%%%%%%%%%
%%%%%%%%%%%%%%%%%%%%%%%%%%%%%%%%%%%%%%%%%%%%%%%%%%%%%%%%%%%%%%%%%%%%%%%%%%%%%%
%%%%%%%%%%%%                    INTRODUCTION                     %%%%%%%%%%%%%
%%%%%%%%%%%%%%%%%%%%%%%%%%%%%%%%%%%%%%%%%%%%%%%%%%%%%%%%%%%%%%%%%%%%%%%%%%%%%%
%%%%%%%%%%%%%%%%%%%%%%%%%%%%%%%%%%%%%%%%%%%%%%%%%%%%%%%%%%%%%%%%%%%%%%%%%%%%%%

\section{Introduction}
In this letter we revisit the definition of the probability
current for the Schrödinger equation. We discuss the probability
current of the hydrogen atom and the isotrop harmonic oscillator
for which the classical results can be expressed in the form: "the
probability current of the wave eigenfunctions $\psi_{nlm}$ of
Schrödinger with $m=0$ is nil", cf. for
example~\cite{CohenTannoudji}. Indeed, as the function
$\psi_{nl0}=R_{nl}(r) Y_{l0}(\theta , \varphi)$ is real, the
classical definition of the Schrödinger probability current leads
to
\begin{equation}\label{1}
  {\bf J_1}=\frac{i \hbar}{2m}(\psi \nabla\psi^* - \psi^* \nabla\psi)=0.
\end{equation}

First, we prove that for the Dirac equation, the probability
current of stationary wave functions of the hydrogen atom and of
the isotrop harmonic oscillator are not nil and correspond to a
circular rotation of the probability.

Then, we recall how a good approximation of the Dirac probability
current is obtained in the nonrelativist case. It is necessary to
add in equation~(\ref{1}) the Gordon current~\cite{Gordon28-1}
which reads
\begin{equation}\nonumber
    {\bf J_{2}}=\frac{\hbar}{2m}rot(\psi^* \sigma \psi)
\end{equation}
in the Pauli approximation~\cite{Gurtler},~\cite{Landau89-1} and
\begin{equation}\label{3}
    {\bf J_2}=\frac{1}{m}\nabla \rho \times \bf s
\end{equation}
in the Schrödinger approximation~\cite{Holland99},~\cite{Colijn},
where $\bf s$ corresponds at a constant spin vector. Consequently,
we get a circular probability current in the Schrödinger
approximation for the hydrogen atom and the isotrop harmonic
oscillator.

%%%%%%%%%%%%%%%%%%%%%%%%%%%%%%%%%%%%%%%%%%%%%%%%%%%%%%%%%%%%%%%%%%%%%%%%%%%%%%
%%%%%%%%%%%%%%%%%%%%%%%%%%%%%%%%%%%%%%%%%%%%%%%%%%%%%%%%%%%%%%%%%%%%%%%%%%%%%%
%%%%%%%%%%%% The probability flux in the Dirac equation            %%%%%%%%%%
%%%%%%%%%%%%%%%%%%%%%%%%%%%%%%%%%%%%%%%%%%%%%%%%%%%%%%%%%%%%%%%%%%%%%%%%%%%%%%
%%%%%%%%%%%%%%%%%%%%%%%%%%%%%%%%%%%%%%%%%%%%%%%%%%%%%%%%%%%%%%%%%%%%%%%%%%%%%%
\section{The Dirac probability current of the hydrogen atom  }

In the presence of electromagnetic couplings, the Dirac equation
reads
\begin{equation}\label{4}
  -i\gamma^{\mu}\frac{\partial \psi}{\partial x_{\mu}}+\frac{e}{\hbar}\gamma^{\mu}A_{\mu} \psi +\frac{mc}{\hbar}\psi=0
\end{equation}
where the $\gamma^{\mu}$ ($\mu=0,1,2,3$) are the Dirac matrices.

The probability density $\rho =\psi^\dag\psi$ and the three
components of the probability current $ {\bf J} $:
\begin{equation}\label{diraccurrent}
  J^{i}=c \psi^{\dag} \gamma^0 \gamma^{i} \psi
\end{equation}
satisfy the continuity equation
\begin{equation}\label{eqcon}
  \frac{\partial \rho}{\partial t} +{\nabla}.{\bf J}=0.
\end{equation}

In the case of an electron in the hydrogen atom with a spherically
symmetric potential $V=-\frac{e}{4 \pi r}$ and a potential vector
$ {\bf A} =0$, the Dirac eigenvalues equation is then
\begin{equation*}
{\gamma_0}({-\frac{E}{{\hbar} c}+\frac{e V}{{\hbar} c}){\psi (\bf
r)}-i{\bf \Gamma}.{\nabla}{\psi (\bf r)} +\frac{mc}{\hbar}\psi(\bf
r)=0}.
\end{equation*}

The eigenfunctions $\psi _{n,l,j,m}(\bf r)$ depend on the quantum
numbers $n,l,j$ and $m$ with $n$ integer ($n \ge 1$), $l$ integer
($0 \le
 l \le n-1$), $j=\frac{1}{2}$ if $l=0$ and $j=l+m$ if $l \ge 1$,
 $m=\pm \frac{1}{2}$. Then $\psi _{n,l,j,m}(\bf r)$ is given~\cite{Landau89-1} by
\begin{equation}\nonumber
\psi _{n,l,j,m}({\bf r})=\begin{pmatrix}
  f(r) \Omega_{jlm} \\
  ig(r) \Omega_{jl'm}
\end{pmatrix}
\end{equation}
with $l'=j+\frac{1}{2}$ if $l=j-\frac{1}{2}$ and
$l'=j-\frac{1}{2}$ if $l=j+\frac{1}{2}$, and where $f(r)$ and
$g(r)$ are real functions of $r$ and where
\begin{equation}\nonumber
 {\Omega}_{l \pm \frac{1}{2},l,m}=\begin{pmatrix}
   \pm \sqrt{\frac{l+\frac{1}{2}\pm{m}}{2l+1}}Y_{l,m-\frac{1}{2}} \\
   \sqrt{\frac{l+\frac{1}{2}\mp{m}}{2l+1}}Y_{l,m+\frac{1}{2}} \
 \end{pmatrix}
 .
\end{equation}
We obtain four classes of wave eigenfunctions: $m=\frac{1}{2}$ or
$-\frac{1}{2}$, $l=0$ or $l\ge 1$. For $m=\frac{1}{2}$
\begin{equation}\label{16}
\psi _{n,l,j,m=+\frac{1}{2}}(r,\theta,\varphi)=\begin{pmatrix}
  f(r)\begin{pmatrix}
    a(\theta) \\
     b(\theta)\exp(i
\varphi) \
  \end{pmatrix} \\
  ig(r)\begin{pmatrix}
    c(\theta) \\
    d(\theta)\exp(i \varphi) \
  \end{pmatrix}
\end{pmatrix}
\end{equation}
with for $l=0$: $a(\theta)=Y_{00}$, $b(\theta)=0$,
$c(\theta)=-Y_{00}\cos\theta$, $d(\theta)=-Y_{00}\sin\theta$ and,
for $l>0$: $a(\theta)=\sqrt{\frac{l+1}{2l+1}}Y_{l,0}$,
$b(\theta)=\sqrt{\frac{l}{2l+1}}Y_{l,1}\exp(-i \varphi)$,
$c(\theta)= \sqrt{\frac{l+1}{2l+3}}Y_{l+1,0} $ and
$d(\theta)=\sqrt{\frac{l+2}{2l+3}}Y_{l+1,1}\exp(-i \varphi)$, and
for $m=-\frac{1}{2}$
\begin{equation}\label{17}
\psi _{n,l,j,m=-\frac{1}{2}}(r,\theta,\varphi)=\begin{pmatrix}
  f(r)\begin{pmatrix}
    -b(\theta) \exp(-i \varphi) \\
     a(\theta) \
  \end{pmatrix} \\
  ig(r)\begin{pmatrix}
    d(\theta) \exp(-i \varphi)\\
    -c(\theta) \
  \end{pmatrix}
\end{pmatrix}
\end{equation}
with for $l=0$: $a(\theta)=0$, $b(\theta)=Y_{00}$,
$c(\theta)=-Y_{00}\sin\theta$, $d(\theta)=Y_{00}\cos\theta$ and,
for $l>0$: $a(\theta)=-\sqrt{\frac{l+1}{2l+1}}Y_{l,-1}\exp(i
\varphi)$, $b(\theta)=\sqrt{\frac{l}{2l+1}}Y_{l,0}$, $c(\theta)=
-\sqrt{\frac{l-1}{2l-1}}Y_{l-1,-1}\exp(i \varphi) $ and
$d(\theta)=\sqrt{\frac{l}{2l-1}}Y_{l-1,0}$.

For a wave function $\psi=\begin{pmatrix}
  \chi_{1} \\
  \chi_{2}
\end{pmatrix}$
and the Dirac matrices
\begin{equation*}
\gamma^0=\begin{pmatrix}
  I & 0 \\
  0 & -I
\end{pmatrix}
\qquad  and \qquad   \gamma^i=\begin{pmatrix}
  0 & \sigma^i \\
  -\sigma^i & 0
\end{pmatrix}
\end{equation*}
the probability density is equal to: $ \rho={\bf \chi}_1^*{\bf
\chi}_1+{\bf \chi}_2^*{\bf \chi}_2$, and the probability current
is given by: $ J^i =c({\bf \chi}_1^* {\sigma^i} {\bf \chi}_2+{\bf
\chi}_2^* \sigma^i {\bf \chi}_1)$.

For the functions~(\ref{16}) and~(\ref{17}), we find~: $
    \rho=f^2(r)[a^2(\theta)+b^2(\theta)]+g^2(r)[c^2(\theta)+d^2(\theta)]$,
     $J^1 =-2cf(r)g(r)[ad-bc]\sin \varphi$, $
    J^2 =-2cf(r)g(r)[ad-bc]\cos \varphi$, $
    J^3 =0$,
and as probability current velocity:
\begin{equation}\label{20}
 {\bf v}=\frac{\bf J}{\rho} =\frac{2cf(r)g(r)[ad-bc]}
 {f^2(r)[a^2+b^2]+g^2(r)[c^2+d^2]}{\bf
 u_\varphi}
\end{equation}
which corresponds to circular probability currents.

We have the same result~(\ref{20}) for all the Dirac equations
with a spherically symetric potential; in particular for the
isotrop harmonic oscillator.

The Dirac eigenfunctions are not stationary. All the electron wave
eigenfunctions of the hydrogen atom correspond to circular
probability currents. It is not the case for the Schrödinger
equation with the current given by~(\ref{1}), since we have then
stationary solutions for the eigenfunctions $\psi_{nl0}$. In the
fondamental state $1s_{\frac{1}{2}}$ with the energy
$E_{n,\frac{1}{2}}=m_e c^2 \sqrt{1-\alpha^2}$, we have: $
    f(r)=a_0(\frac{r}{r_0})^{\sqrt{1-\alpha^2}-1}e^{-\frac{r}{r_0}}$,
    $
    g(r)=- a_0
    \frac{1-\sqrt{1-\alpha^2}}{\alpha}(\frac{r}{r_0})^{\sqrt{1-\alpha^2}-1}e^{-\frac{r}{r_0}}$,
where $r_0=\frac{\hbar^2}{m_e e^2}=0.53~\AA$ is the Bohr radius.
From~(\ref{20}), we deduce $
    \rho(r,\theta,\varphi)=\rho(r)
    =2 a_0^2\frac{1-\sqrt{1-\alpha^2}}{\alpha^2}(\frac{r}{r_0})^{2(\sqrt{1-\alpha^2}-1)}
    e^{-2\frac{r}{r_0}}$, and
\begin{equation}\label{22}
{\bf v}=\alpha c \sin \theta {\bf u_{\varphi}}.
\end{equation}
From~(\ref{17}) and~(\ref{20}), we deduce ${\bf v}=-\alpha c \sin
\theta {\bf u_{\varphi}}$.
 There is a big difference between~(\ref{22}) and the current nil
given by~(\ref{1}) with the Schrödinger equation. The lesson of
this example is that, although Schrödinger equation is a good
approximation of the Dirac equation, the Schrödinger probability
current~(\ref{1}) is not the good approximation of the Dirac
probability current~(\ref{diraccurrent}). This conclusion appears
clearly, as we recall
now~\cite{Gordon28-1},~\cite{Gurtler},~\cite{Landau89-1}, in the
approximations to transform first the bispinor of the Dirac
equation to the spinor of the Pauli equation, then the Pauli
spinor to the wave function of the Schrödinger equation.

%%%%%%%%%%%%%%%%%%%%%%%%%%%%%%%%%%%%%%%%%%%%%%%%%%%%%%%%%%%%%%%%%%%%%%%%%%%%%%
%%%%%%%%%%%%%%%%%%%%%%%%%%%%%%%%%%%%%%%%%%%%%%%%%%%%%%%%%%%%%%%%%%%%%%%%%%%%%%
%%%%%%%%%%%% The probability current in the Pauli and Schrödinger equation%%%%%%%%
%%%%%%%%%%%%%%%%%%%%%%%%%%%%%%%%%%%%%%%%%%%%%%%%%%%%%%%%%%%%%%%%%%%%%%%%%%%%%%
%%%%%%%%%%%%%%%%%%%%%%%%%%%%%%%%%%%%%%%%%%%%%%%%%%%%%%%%%%%%%%%%%%%%%%%%%%%%%%
\section{The Pauli and Schrödinger probability currents  }

 After the change of variable $
    \psi=\psi' e^{-\frac{i}{\hbar} mc^2t}$,
the Dirac equation~(\ref{4}) for the bispinor $\psi=(
\chi_1,\chi_2) $ can be written
\begin{equation}\label{23}
    [i \hbar \frac{\partial}{\partial t}-eV] \chi_1
    =c {\bf \sigma}(-i\hbar \nabla - \frac{e}{c} {\bf A}) \chi_2
\end{equation}
\begin{equation}\label{24}
    [i \hbar \frac{\partial}{\partial t}-eV+ 2m c^2] \chi_2
    =c {\bf{\sigma}}(-i \hbar \nabla - \frac{e}{c} {\bf A}) \chi_1.
\end{equation}
Small electron velocities allow us to neglect the two first terms
of the left part of equation~(\ref{24}) and to eliminate $\chi_2$
in (\ref{23}). So we obtain the Pauli equation for the spinor
$\chi_1$
\begin{equation}\nonumber
    i \hbar \frac{\partial \chi_1}{\partial t}
    =[\frac{1}{2m}(-i\hbar \nabla - \frac{e}{c} {\bf A})^2 +eV - \frac{e \hbar}{2m c}
    {\bf \sigma}.{\bf H}]\chi_1
\end{equation}
where $\bf H=\nabla \times A$ is the magnetic field.

As $\chi_2 \ll \chi_1$, the Dirac density $\rho=\psi'^* \psi=
\chi_1^* \chi_1 +\chi_2^* \chi_2$ is equal in first approximation
to the density $\rho=\chi_1^* \chi_1$ of the Pauli equation. Then
the Dirac current density can be written
\begin{equation}\label{28}
{\bf J}=i\frac{\hbar}{2m}(\chi_1{\bf \nabla}\chi_1^* -\chi_1^*{\bf
\nabla}\chi_1 )-\frac{e}{mc} {\bf A}
\chi_1\chi_1^*+\frac{\hbar}{2m} \nabla \times (\chi_1^*{\bf
\sigma}\chi_1)
\end{equation}
Thus, to obtain in Pauli a good approximation of the Dirac
probability current, it is necessary to add the Gordan current
~\cite{Gordon28-1}
\begin{equation}\nonumber
    {\bf J_2}
    =\frac{\hbar}{2m} \nabla \times (\chi_1^*{\bf\sigma}\chi_1)
\end{equation}
to the classical Pauli current
\begin{equation}\nonumber
    {\bf J_1}
    =i\frac{\hbar}{2m}(\chi_1{\bf \nabla}\chi_1^*
    -\chi_1^*{\bf \nabla}\chi_1 )-\frac{e}{m c}\bf A \chi_1\chi_1^*.
\end{equation}
Then, the current $\bf J=\bf J_1+\bf J_2$ verifies, as the current
$\bf J_1$, the continuity equation(\ref{eqcon}).

To obtain the Schrödinger equation from Pauli equation, consider
the case where we have no magnetic field~\cite{Holland99} and
where the system is in a spin eigenstate
\begin{equation}\nonumber
\chi_1({\bf r},t)= \varphi({\bf r},t) \chi
\end{equation}
with a constant spinor $\chi$ such as $\chi^* \chi=1$. Then the
function $\varphi({\bf r},t)$ verifies the Schrödinger equation
\begin{equation}\label{32}
    i \hbar \frac{d \varphi}{dt}
    =- \frac{\hbar^2}{2m} \Delta\varphi +e V \varphi .
\end{equation}

However, in this case, the current~(\ref{28}) does not
reduce~\cite{Holland99} to classical Schrödinger probability
current. Indeed, writing $\varphi({\bf r},t)=\sqrt{\rho}
e^{i\frac{S}{\hbar}}$, we have
\begin{equation}\nonumber
    {\bf J}
    =\frac{\rho}{m} \nabla S +\frac{1}{m}\nabla \rho \times \bf s
\end{equation}
with
\begin{equation}\nonumber
    {\bf s}=\frac{\hbar}{2} (\chi^*{\bf \sigma}\chi).
\end{equation}

Indeed, to obtain a good approximation of the Dirac probability
current for the Schrödinger equation, it is necessary to add the
spin-dependent current
\begin{equation*}
    {\bf J_2}=\frac{1}{m}\nabla \rho \times \bf s
\end{equation*}
to the classical Schrödinger current
\begin{equation*}
    {\bf J_1}=\frac{\rho}{m} \nabla S .
\end{equation*}

Therefore the probability current depends on a constant spin
vector $\bf s$. The add of a constant spin vector to the
Schrödinger equation is not new, and has been used before by
Landau~\cite{Landau89-2}.

This current $\bf J_2$ gives also a simple explanation of the
value 2 of the gyromagnetic ratio of the Dirac electron; Indeed,
this contribution to the particle orbital angular momentum is
\begin{equation*}
    \textbf{L}_\textbf{2}=\textbf{r}\times m \textbf{J}_\textbf{2}=\textbf{r} \times
\nabla\rho \times \textbf{s}
\end{equation*}
and this mean angular momentum is then
\begin{equation*}
   <\textbf{L}_\textbf{2}>=\int \textbf{L}_\textbf{2}d^3r=
   \int (\textbf{r}\cdot\textbf{s})\nabla\rho  d^3r
   -\int(\textbf{r}\cdot \nabla\rho)\textbf{s}  d^3r=-\textbf{s}+3\textbf{s}= 2\textbf{s}.
\end{equation*}

We can verify the precedent theoritical approach with the
computation of the current~(\ref{32}) of the hydrogen atom wave
eigenfunctions in the approximation of Schrödinger.

For the wave functions $
    \psi_{nlm}(r,\theta , \varphi)
    =R_{nl}(r) P_{lm}(\theta)\exp(im\varphi)\exp(-i \frac{E_n
    t}{\hbar})$, the classical Schrödinger current is
\begin{equation}\nonumber
    {\bf J_1}
    =\frac{\rho}{m_e} \frac{m \hbar}{r \sin \theta}{\bf u_{\varphi}}
\end{equation}
with $ \rho=R_{nl}^2(r) P_{lm}^2(\theta)$. The current is nil for
the wave functions with $m=0$, in particular for the states $ns$.
If $m \ge 0$, we take as spin vector ${\bf s}=\frac{\hbar}{2} \bf
k$ and ${\bf s}=-\frac{\hbar}{2} \bf k$ if $m < 0$. Then, this
vector can be written
\begin{equation*}
{\bf s}=sgn(m) \frac{}{}(\cos \theta {\bf u_r} -\sin \theta {\bf
u_{\varphi}})
\end{equation*}
and the additional spin-dependent current (Gordon current) $\bf
J_2$ is then equal to
\begin{equation}\nonumber
    {\bf J_2}=- sgn(m) \frac{\hbar}{m_e}(R_{nl} R'_{nl}P^2_{lm}\sin
    \theta + \frac{R^2_{nl}}{r}P_{lm}P'_{lm} \cos \theta){\bf
    u_{\varphi}}.
\end{equation}
Then, the probability current velocity is
\begin{equation}\nonumber
{\bf v}= \frac{\bf J}{\rho}=\frac{\hbar}{m_e}[- sgn(m)(\frac
{R'_{nl}(r)}{R_{nl}(r)}\sin \theta + \frac{P'_{lm}( \theta)}{r
P_{lm}( \theta)} \cos \theta)+\frac{m}{r \sin \theta}]{\bf
u_{\varphi}}
\end{equation}
which corresponds to circular probability currents exactly as for
the Dirac equation~(\ref{20}). In the case of the fondamental
state $1s$, we find again exactly the velocity~(\ref{22}) of the
Dirac probability current
\begin{equation}\nonumber
{\bf v}_{1s}=\alpha c \sin \theta {\bf u_{\varphi}}.
\end{equation}

For the states $2s$, $2p_0$, $2p_1$ and $2p_{-1}$, we obtain for
the current velocity
\begin{equation}\nonumber
{\bf v}_{2s}=\frac{\alpha c}{2}(1+ \frac{1}{1- \frac{r}{2 r_0}})
\sin \theta {\bf u_{\varphi}},
\end{equation}
\begin{equation}\nonumber
{\bf v}_{2p0}={\bf v}_{2p_1}=\frac{\alpha c}{2} \sin \theta {\bf
u_{\varphi}}  \qquad     and    \qquad      {\bf
v}_{2p_{-1}}=-\frac{\alpha c}{2} \sin \theta {\bf u_{\varphi}}.
\end{equation}

%%%%%%%%%%%%%%%%%%%%%%%%%%%%%%%%%%%%%%%%%%%%%%%%%%%%%%%%%%%%%%%%%%%%%%%%%%%%%%
%%%%%%%%%%%%%%%%%%%%%%%%%%%%%%%%%%%%%%%%%%%%%%%%%%%%%%%%%%%%%%%%%%%%%%%%%%%%%%
%%%%%%%%%%%% Conclusion %%%%%%%%%
%%%%%%%%%%%%%%%%%%%%%%%%%%%%%%%%%%%%%%%%%%%%%%%%%%%%%%%%%%%%%%%%%%%%%%%%%%%%%%
%%%%%%%%%%%%%%%%%%%%%%%%%%%%%%%%%%%%%%%%%%%%%%%%%%%%%%%%%%%%%%%%%%%%%%%%%%%%%%
\section{Conclusion }

As we have shown, although the Schrödinger equation is a good
approximation of the Dirac equation, the Schrödinger probability
current~(\ref{1}) is not the good approximation of the Dirac
probability current~(\ref{diraccurrent}). The main conclusion is
that it is necessary to add to the classical Schrödinger current
the spin-dependent current
\begin{equation*}
    {\bf J_2}=\frac{1}{m}\nabla \rho \times \bf s
\end{equation*}
which corresponds to a constant spin vector $\bf s$.

In the case of the isotrop harmonic oscillator with $V({\bf
r})=\frac{1}{2}m \omega^2 {\bf r}^2$, the fondamental wave
function $1s$
\begin{equation*}
\psi_{0,0,0}({\bf r})=(\frac{m \omega}{\pi \hbar})^\frac{3}{4}
\exp (- \frac{m \omega}{2 \hbar}{\bf r}^2)  \exp( - \frac{3 i
\omega t }{2})
\end{equation*}
gives a Schrödinger classical current $\bf J_1$ nil. If we
consider an oscillator with a constant spin ${\bf s}=
\frac{\hbar}{2} \bf k$, then the equation~(\ref{32}) gives a
circular probability current $\bf J$ with the velocity
\begin{equation}\label{41}
{\bf v}_{1s}=\frac{\bf J}{\rho}=- \omega r \sin \theta {\bf
u_{\varphi}}.
\end{equation}
It is the same velocity as in the classical case. For the other
eigenfunctions of the harmonic oscillator, we find also a non nil
probability current, but non circular as in~(\ref{41}).

\end{document}